\documentclass[a4paper,preprint,superscriptaddress,preprintnumbers,nofootinbib]{revtex4-1}
\newcommand{\Slash}[1]{{\ooalign{\hfil/\hfil\crcr$#1$}}}

\usepackage{amsmath}	% required for `\split' (yatex added)
\usepackage{mathrsfs}
\usepackage[dvipdf]{graphicx}
\usepackage{subfigure}

\begin{document}

\title{Analysis of Spontaneous Mass Generation by Iterative Method \\
in the Nambu-Jona-Lasinio Model and Gauge Theories
\footnote{Presented at ``SCGT12 KMI-GCOE Workshop on Strong Coupling Gauge
Theories in the LHC Perspective'', 4-7 Dec. 2012, Nagoya University.}}

\author{Ken-Ichi \surname{Aoki}}
\email{aoki@hep.s.kanazawa-u.ac.jp}
\affiliation{Institute for Theoretical Physics, Kanazawa University, Kanazawa 920-1192, Japan}

\author{Shinnosuke \surname{Onai}}
\email{onai@hep.s.kanazawa-u.ac.jp}
\affiliation{Institute for Theoretical Physics, Kanazawa University, Kanazawa 920-1192, Japan}

\author{Daisuke \surname{Sato}}
\email{satodai@hep.s.kanazawa-u.ac.jp}
\affiliation{Institute for Theoretical Physics, Kanazawa University, Kanazawa 920-1192, Japan}

\preprint{KANAZAWA-13-06}

\begin{abstract}
We propose a new iterative method to directly calculate
the spontaneous mass generation due to the dynamical chiral
symmetry breaking. We can conclude the physical mass definitely without
recourse to any other consideration like the free energy comparison.
\end{abstract}

\maketitle

\section{Introduction}
Owing to its outstanding feature, dynamical chiral symmetry breaking phenomena
have been widely studied in various fields of elementary particle physics. The
standard method to discuss the spontaneous mass generation is to formulate a
coupled system of self-consistent equations and find its non-trivial solution. 
However, those equations are no more than the {\sl necessary} condition and it is
needed to examine solutions to select correct one by using another mean e.g. by
referring to the free energy of each solution. Even if it is done, there are
still unclear points whether the minimal free energy solution ensures the
physically correct answer.

We adopt the Nambu-Jona-Lasinio (NJL) model\cite{nambu} and give a new
iteration method that directly sums up an infinite number of diagrams of the
standard perturbation theory in the ladder approximation. Using this method, we
demonstrate that the physically meaningful result is automatically obtained with
the correct critical coupling constant. The NJL model has four-fermion 
interactions among the massless fermions with the chiral invariance. Here we
added the bare mass $m_0$ to the Lagrangian to make the standard perturbation
theory work well,
\begin{equation}
 {\cal{L}}_{\rm{NJL}} = \bar{\psi}\Slash{\partial}\psi+
 \frac{2{\pi}^2g}{N} \left[\left({\bar{\psi}\psi}\right)^2 + 
 \left(\bar{\psi}i\gamma_{5}\psi\right)^2\right] - m_0\bar{\psi}\psi,
\end{equation}
where $N$ is the number of fermion flavors. We consider $1/N$ leading
contribution to the mass. Diagrammatically it is a sum of infinite diagrams
called `tree', where considering the fermion-antifermion pair 
as a single meson, the tree diagrams are defined by those without any 
meson loops, or in other words we regard a series of loops as a {\sl fat}
propagator.Usual method is to set up an equation satisfied by this infinite 
sum of diagrams. Above the critical coupling constant, there actually exists a
non-trivial solution which does not vanish after zero bare mass limit. However,
it is unclear that how the finite mass should come out of the infinite sum of
the diagrams while each finite diagram must vanish at the zero bare mass limit.
In this article we set up a method to directly sum up the infinite diagrams and
show how the finite mass come out without any ambiguity\cite{onai13}.

\section{Iteration Method}
First of all we classify diagrams in the {\sl tree} using the node length of each
diagram. Node length of a diagram is defined by the maximum number of loops in a
continuous route towards the edge loop, or maximum number of nodes of {\sl fat}
propagator legs in the diagram. Fig.~1(a) shows the counting rule of node length
and classification of diagrams. Here we define $M^{(n)}$ is a sum of diagrams
whose node length is no greater than $n$.

\begin{figure}[h!]
 \begin{center}
 \subfigure[Definition.]{
 \includegraphics[width=2.4in]{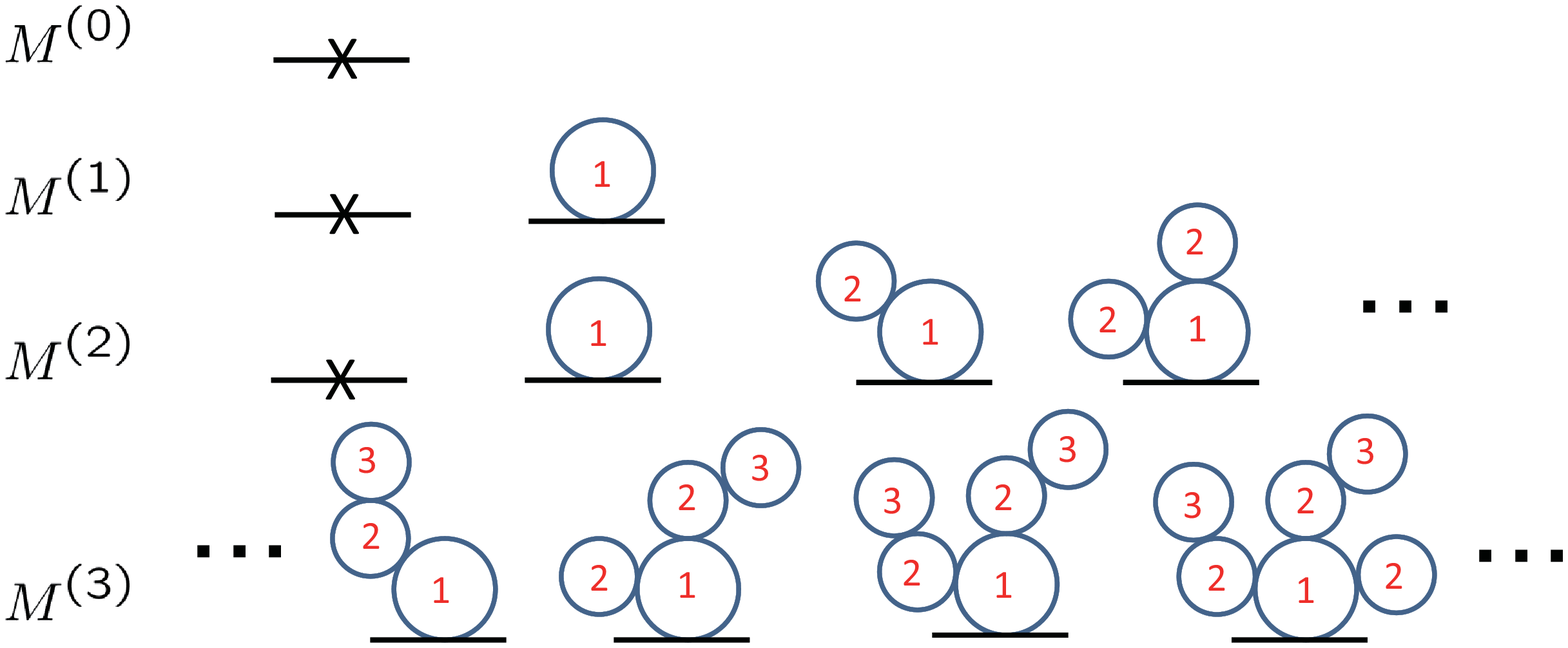}}
 \subfigure[Node length iteration.]{
 \includegraphics[width=2.4in]{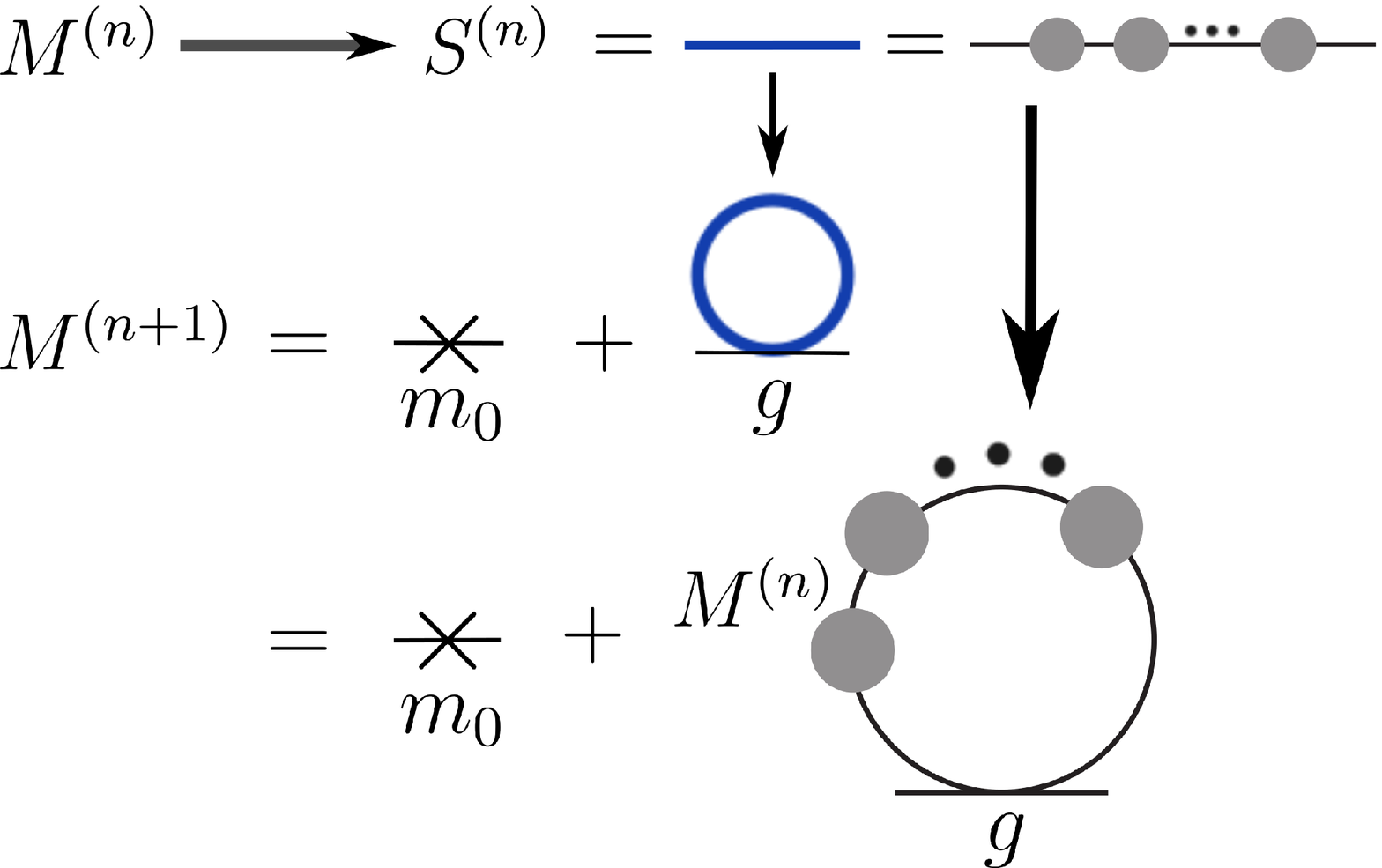}}
 \caption{\label{fig:qm/complexfunctions} Node length.}
 \end{center}
 \vskip-0.5cm
\end{figure}

Then we write down the iteration to calculate $M^{(n+1)}$ by using $M^{(n)}$ as
shown in Fig.~1(b). The transformation function is a one loop integral and we 
denote it by $F$,
\begin{equation}
 M^{(n+1)}=F\left(M^{(n)}\right),\ \ 
 F(M)=m_0+gM\left(1-M^2\log\left(1+M^{-2}\right)\right).
\end{equation}
Therefore the total sum of the tree diagrams is obtained by
$M^{(\infty)}$, infinitely many times of transformation of the same $F$.

\section{Mass Generation}
Iterative transformation here is best understood by a graphical method where the
transformation function $y=F(x)$ and a straight line of $y=x$ are drawn as shown
in Fig.~2. Each iteration process can be drawn on this figure by a successive
move of points. In any case the iterative transformation finally reaches a
stable fixed point. Fixed points are crossing points between $y=F(x)$ and $y=x$,
and position of fixed points are shown in Fig.~3.

\begin{figure}[h!]
 \begin{center}
 \subfigure[Weak coupling : $g$=0.7.]{
 \includegraphics[width=2.4in]{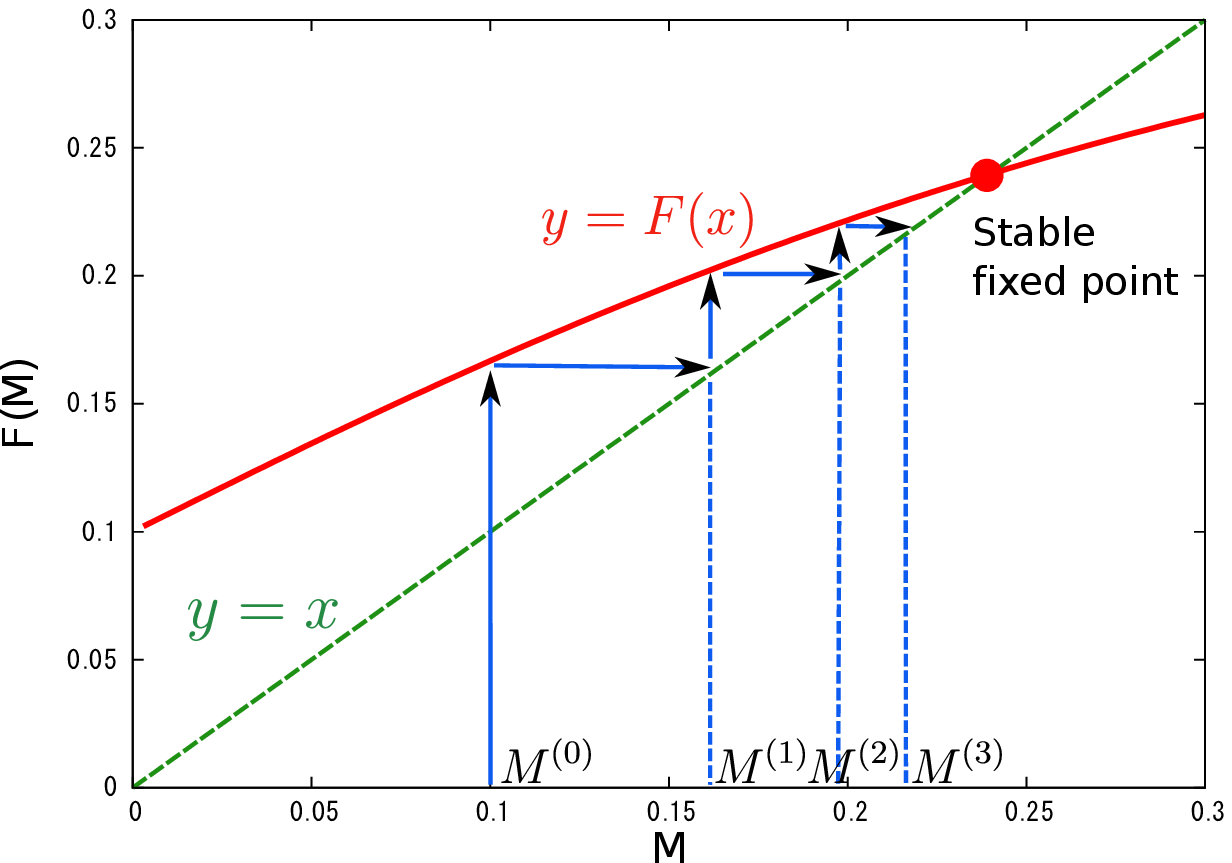}}
 \subfigure[Strong coupling : $g$=1.5.]{
 \includegraphics[width=2.4in]{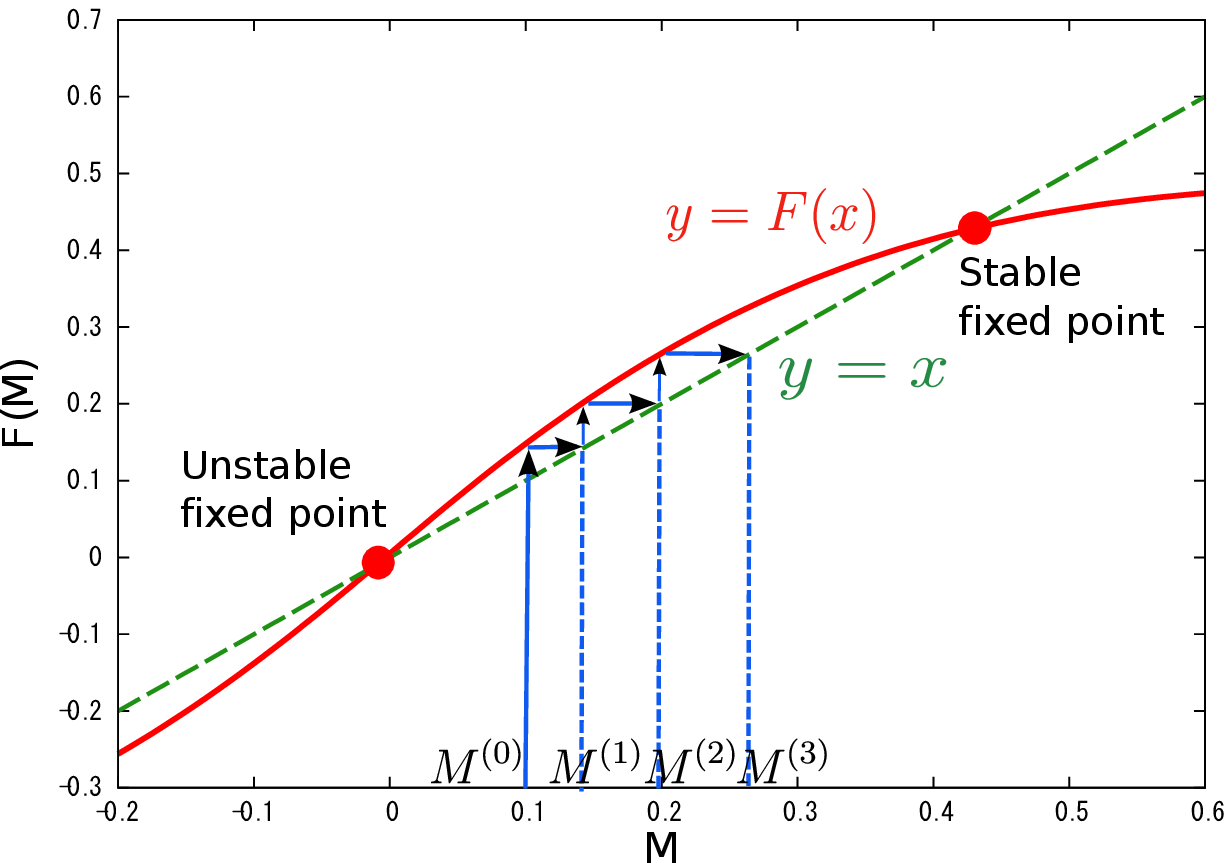}}
 \caption{Iterative steps.}
 \end{center}
 \vskip-0.5cm
\end{figure}

In the weak coupling region, there is only one fixed point near the origin 
which is stable, and iteration from any starting point must reach it, as shown
in Fig.~2(a). When the coupling constant becomes strong, there appear pair
creation of fixed points, one is stable and the other is unstable. Then there
are two stable fixed points each of which has its attractive region, `territory'. 
We must be careful about the initial starting point of iteration,
that is, $M^{(0)}$, which must be the bare mass $m_0$ by definition. In Fig.~3
we set a positive value for the bare mass, then the initial point is in the
territory of the right-hand side stable fixed point, as seen in Fig.~2(b).
Therefore for all region of the coupling constant, the physical result is
controlled by the right most stable fixed point in Fig.~3. The critical coupling
constant for the change of fixed point structure depends on the bare mass and it
becomes unity in the vanishing bare mass limit.

\begin{figure}[htbp]
 \centering
 \includegraphics[width=3.5in]{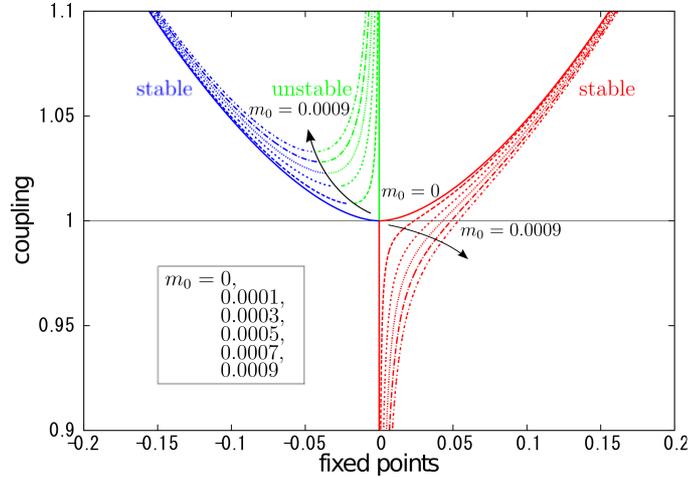}
 \caption{Fixed point structure with the coupling constant and the bare mass.}
 \vskip-0.5cm
\end{figure}

Let's see some features of mass generation with respect to the node length $n$ in
Fig.~4. In the weak coupling case (a), the dynamical mass is generated rather
quickly at low $n$ and becomes constant, which should be called the perturbative
characteristics. By decreasing the bare mass the final mass goes to zero. In the
strong coupling case (b), the generation of the dynamical mass depends strongly
on the bare mass, and it is mainly generated at some narrow range of node 
length. Decreasing the bare mass, the region of mass generating node length
becomes large, but the output mass is almost constant, which means the
spontaneous mass generation. It is also seen that the shape of generation curves
look the same form, just displacement in the node length space. These features
are readily verified by the iterative nature of our calculation well seen in
Fig.~2(b).

\begin{figure}[h!]
 \begin{center}
 \subfigure[Weak coupling : $g$=0.9.]{
 \includegraphics[width=2.4in]{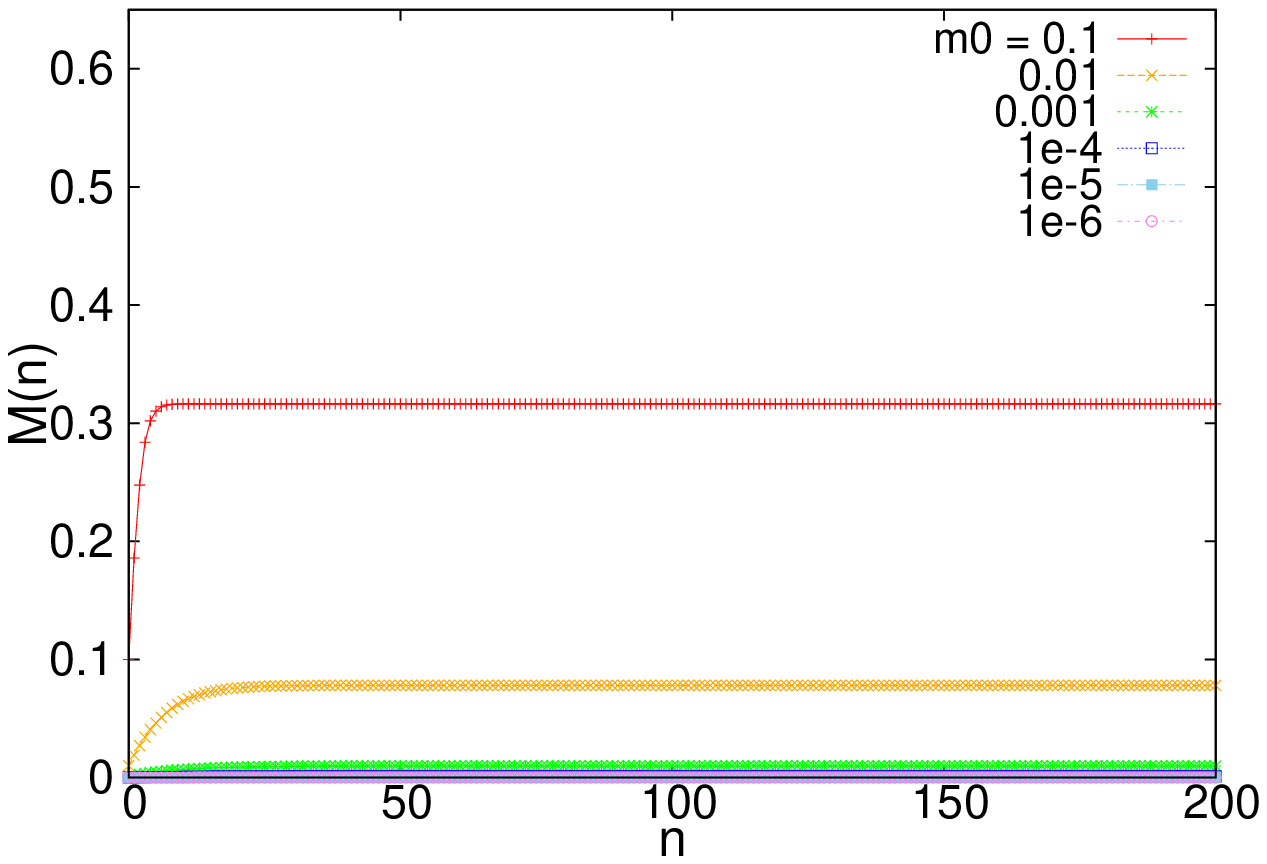}}
 \subfigure[Strong coupling : $g$=1.1.]{
 \includegraphics[width=2.4in]{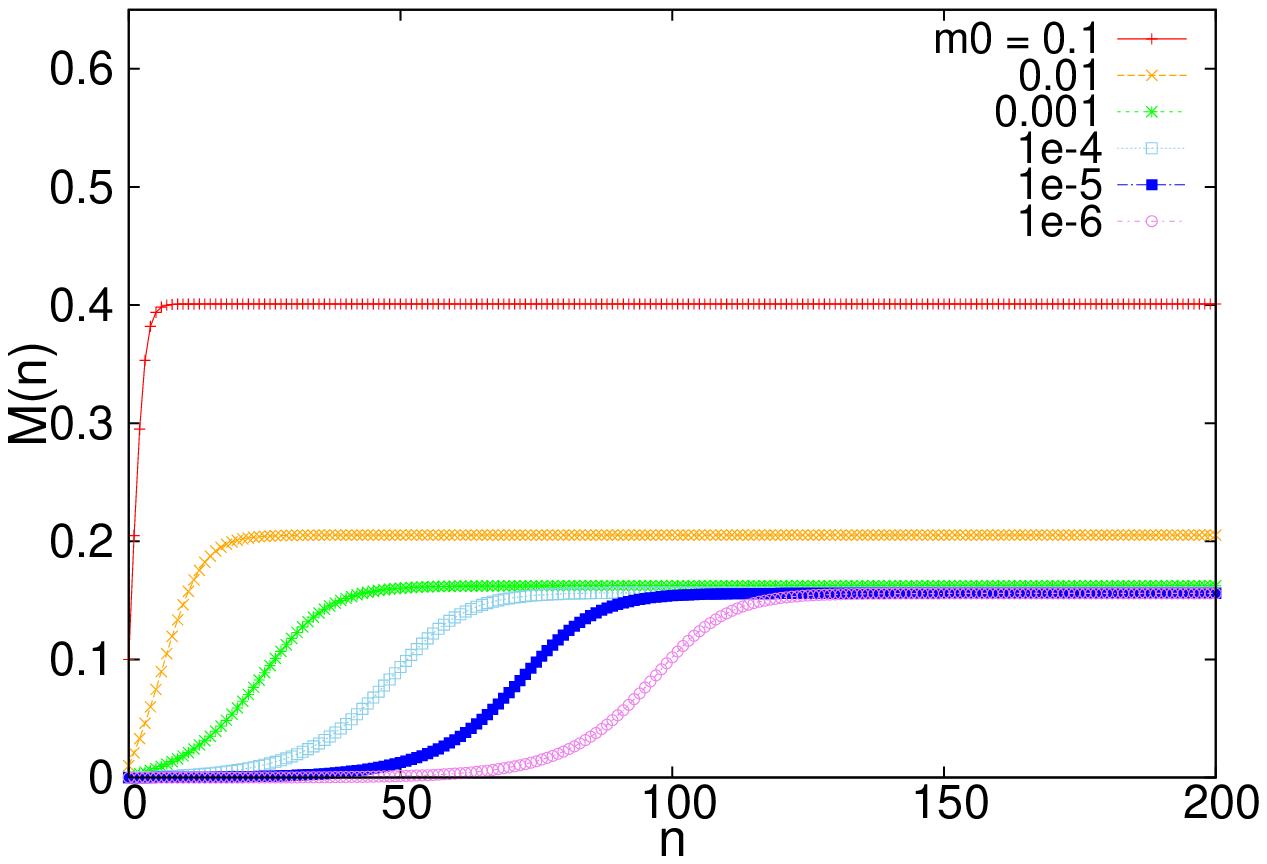}}
 \caption{Mass generation by the node length iteration.}
 \end{center}
 \vskip-0.5cm
\end{figure}

Finally we mention about gauge theories. We add all planar diagrams
which are equivalent to the ladder Schwinger-Dyson equation\cite{aoki}.
To set up iterative method, we define ladder depth of each planar 
diagram which is the maximum number of gluon propagators towards
the fermion propagator. We define mass function
$\Sigma^{(n)}$ which contains all planar diagrams whose ladder depth
is no greater than $n$. Then we write down the iteration transformation
as follows.
\begin{eqnarray}
 {\Sigma}^{(n+1)}(x)&=&\mathscr{F}\left[{\Sigma}^{(n)}(x)\right], \\
 \mathscr{F}\left[\Sigma(x)\right] &=& m_0 +
 \frac{\lambda(x)}{4x}\int_{0}^{x}\frac{y\Sigma(y)dy}{y+\Sigma^2(y)} +
 \int_{x}^{\Lambda^2}\frac{\lambda(y)\Sigma(y)dy}{4(y+\Sigma^2(y))},
\end{eqnarray}
where the mass function $\Sigma(x)$ is a function of momentum squared $x$.
The functional $\mathscr{F}$ is now an infinite dimensional map and
there are infinite number of fixed point functions. Our analysis
clarified that only one of fixed points is perfectly stable and 
reached by proper initial function $\Sigma(x)=m_0$\cite{onai13}.

\end{document}